\documentclass{Interspeech2024}

\interspeechcameraready

\title{DualVC 3: Leveraging Language Model Generated Pseudo Context for End-to-end Low Latency Streaming Voice Conversion\vspace{-25pt}
}

\name[affiliation={1,2}]{Ziqian}{Ning}
\name[affiliation={3}]{Shuai}{Wang}
\name[affiliation={2}]{Pengcheng}{Zhu}
\name[affiliation={1}]{Zhichao}{Wang}
\name[affiliation={1}]{Jixun}{Yao}
\name[affiliation={1*}]{Lei}{Xie}
\name[affiliation={2}]{Mengxiao}{Bi}

\address{
\vspace{-9pt}
  $^1$Audio, Speech and Language Processing Group (ASLP@NPU), School of Computer Science, \\ Northwestern Polytechnical University, Xi'an, China\\
  $^2$Fuxi AI Lab, NetEase Inc., Hangzhou, China \\
  $^3$Shenzhen Research Institute of Big Data, \\The Chinese University of Hong Kong, Shenzhen (CUHK-Shenzhen), China\vspace{-3pt}}
\email{\{ningziqian, zcwang\_aslp, yaojx\}@mail.nwpu.edu.cn, wsstriving@gmail.com, \{zhupengcheng, bimengxiao\}@corp.netease.com, lxie@nwpu.edu.cn\thanks{* Corresponding author.}}


\begin{document}

\maketitle

\begin{abstract}

Streaming voice conversion has become increasingly popular for its potential in real-time applications. The recently proposed DualVC 2 has achieved robust and high-quality streaming voice conversion with a latency of about 180ms. Nonetheless, the recognition-synthesis framework hinders end-to-end optimization, and the instability of automatic speech recognition (ASR) model with short chunks makes it challenging to further reduce latency. To address these issues, we propose an end-to-end model, DualVC 3. With speaker-independent semantic tokens to guide the training of the content encoder, the dependency on ASR is removed and the model can operate under extremely small chunks, with cascading errors eliminated. A language model is trained on the content encoder output to produce pseudo context by iteratively predicting future frames, providing more contextual information for the decoder to improve conversion quality. Experimental results demonstrate that DualVC 3 achieves comparable performance to DualVC 2 in subjective and objective metrics, with a latency of only 50 ms. We have made our audio samples publicly available.~\footnote{Demo: https://nzqian.github.io/dualvc3/}
\end{abstract}

\section{Introduction}
Voice conversion (VC) is a technique that changes the timbre from one speaker to another without altering the semantic information~\cite{vcoverview}. With the development in deep learning, advanced VC models have reached a level of naturalness indistinguishable from actual human speech while maintaining a high degree of speaker similarity. These models have been successfully applied to numerous scenarios such as movie dubbing~\cite{bgmvc,expressive-vc}, privacy protection~\cite{anonymization}, and pronunciation correction~\cite{ELECTROLARYNGEAL}. Typical VC models~\cite{freevc,acevc,vqmivc,stylevc} accept an entire utterance as input and generate the converted speech as a whole. However, there is an increasing demand for real-time communication (RTC) applications, including live broadcasting and online meetings, requiring the speech to be converted on the fly, which poses a challenge for conventional VC models.

Unlike non-streaming models, streaming models process speech input in frames or chunks, and operate causally without access or access to very limited future information. The absence of future context results in degraded performance. To mitigate this issue, a common approach is to employ knowledge distillation, where a pre-trained non-streaming teacher model~\cite{ibfvc} or non-streaming path built in the model~\cite{dualvc,dualvc2} provides additional guidance. Implicit knowledge distillation is used in~\cite{DBLP:conf/icassp/HayashiKT22} by sharing convolutional parameters, using the full convolutional receptive field in non-streaming inference, and excluding receptive field involving future information in streaming inference. Additionally, in ~\cite{ibfvc}, intermediate bottleneck features from the middle layers of the ASR encoder are leveraged, which preserves more information to compensate for mispronunciations caused by streaming ASR's degraded performance. 
Another possible approach is to predict pseudo future context, for instance, CUSIDE~\cite{cuside} designs a simple feed-forward layer to simulate the future context, thereby improving the performance of streaming ASR. This approach shares a conceptual similarity with language modeling, which is trained by predicting future steps.
 
\begin{figure*}[!htbp]
\centering
\includegraphics[scale=0.42]{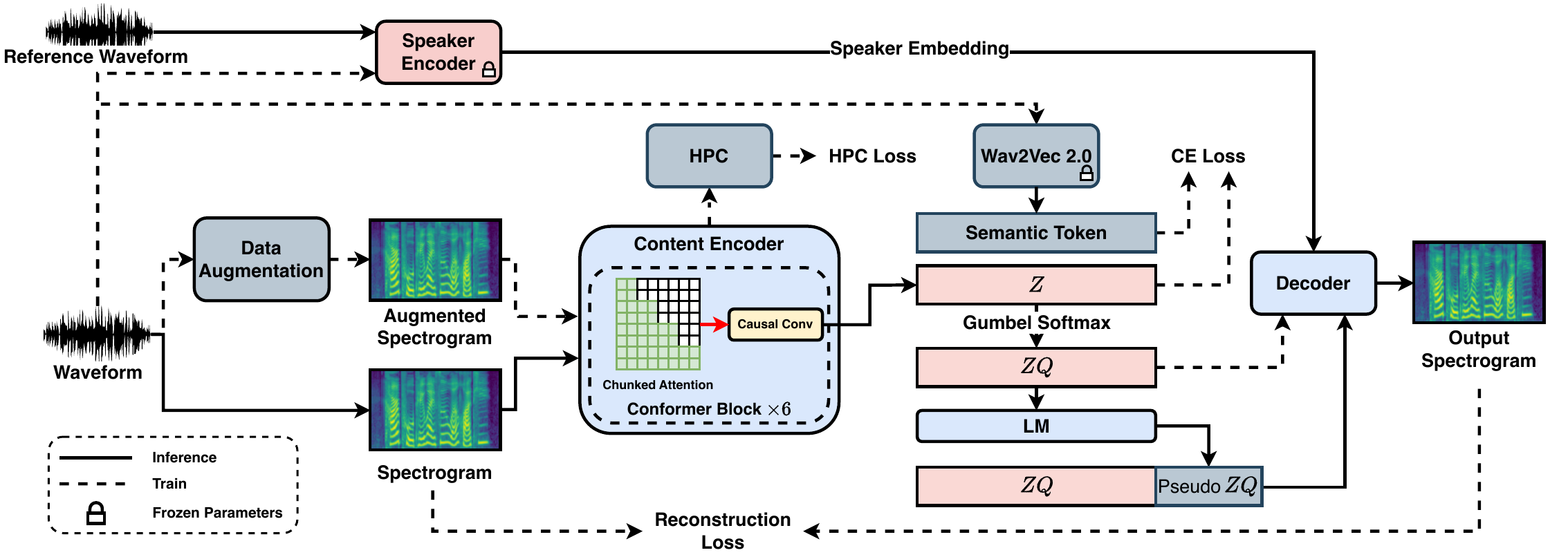}
\vspace{-6pt}
\caption{The architecture of DualVC 3. Implicit knowledge is achieved by applying a dynamic chunk mask in Conformer blocks, allowing dual-mode inference that can accept both full sequence and chunked speech as input. HPC and Wav2Vec 2.0 are discarded after training.}
\label{fig:model}
\vspace{-14pt}
\end{figure*}


At present, most high-quality streaming voice conversion models~\cite{ibfvc,dualvc,dualvc2,streamvoice} are based on the recognition-synthesis framework. This framework benefits from ASR models trained on extensive amounts of lossy data, allowing for the extraction of robust semantic information.
However, such recognition-synthesis framework has several limitations in streaming conversion scenarios. 
First, the multi-level cascaded models, consisting of an ASR encoder, an acoustic model, and a vocoder, can cause inevitable cascading errors.
Deploying the cascaded models for practical usage can be challenging due to the complexity of the whole pipeline.
Second, streaming ASR models require larger data chunks to achieve optimal performance, restricting the downstream streaming VC models to further reduce latency by shrinking the size of data chunks.
Although unsupervised speech representation disentanglement (SRD) based streaming VC models~\cite{streamingvqmivc, fasts2svc} can get rid of the ASR module, 
they require complex and meticulous model design and are prone to speaker timbre leakage problems.

In this paper, we present \textit{DualVC 3}, a high-quality end-to-end streaming voice conversion model that aims to achieve extremely low latency. 
Instead of using ASR to extract semantic information, we train the content encoder with the guidance of a pre-trained semantic token extractor, Wav2Vec 2.0~\cite{w2v2}, which is not needed in the inference stage, eliminating cascading errors and reducing delays. 
To mitigate the impact of absent future context, we employ a language model to generate pseudo context for the decoder. 
Similar to the recently proposed token based language model~\cite{audiolm, musiclm}, which exhibit powerful in-context learning ability,
we train the language model on the discrete intermediate representations extracted by the content encoder, and iteratively predicts a few frames into the future as an additional context for the decoder in the inference stage.
Following the concept of dual-mode training in~\cite{dualvc,dualvc2}, DualVC 3 is built on DualVC 2~\cite{dualvc2}, with conformer-based encoder and decoder trained using dynamic chunk masks. With the chunk size varying from 1 to full sequence, implicit knowledge distillation is achieved within the model, allowing it to be applied to both streaming and non-streaming scenarios. 
Additionally, DualVC 3 incorporates the HPC module and quiet attention introduced in~\cite{dualvc, dualvc2}, and data augmentation is applied to further improve the model's robustness and intelligibility.
Through extensive experiments, our proposed end-to-end DualVC 3 achieves an extremely low latency of only 50 ms on a single-core CPU, with minimal quality degradation from previous cascaded streaming systems.

\section{Proposed Approach}

\subsection{System Architecture}
DualVC 3 is an end-to-end streaming voice conversion model with the mel-spectrogram as the input and output. It consists of a content encoder, a decoder, which we refer to as the acoustic model (AM), and a language model (LM).

\noindent\textbf{Content Encoder}
The content encoder consists of multiple conformer blocks stacked on top of each other, which takes the mel-spectrogram as input and extracts speaker-independent semantic information. The semantic tokens obtained from K-means clustering of SSL representations is used to perform semantic distillation for the content encoder. 
We apply dynamic chunk training (DCT)~\cite{wenet} to make the conformer streamable.
The concept of DCT involves dynamically varying the chunk size by applying a dynamic chunk mask to the attention score matrix for each self-attention layer. The semantic information extracted by the encoder is further discretized for the language model and decoder.

\noindent\textbf{Language Model}
A language model is trained on the discrete semantic information in the typical next-token-prediction manner. It can iteratively generate pseudo context for the decoder for better conversion quality during inference.

\noindent\textbf{Decoder}
The decoder has the same structure as the content encoder. With the concatenation of embedded discrete semantic information and a global speaker embedding extracted by a pre-trained speaker encoder as input, it generates the mel-spectrogram with converted speaker timbre.

\noindent\textbf{Hybrid predictive coding}
Hybrid predictive coding (HPC) is an unsupervised representation learning method proposed in DualVC~\cite{dualvc}, which is a combination of CPC~\cite{cpc} and APC~\cite{apc}. Here we compute the HPC loss on the intermediate representations of the content encoder, enhancing the encoder's contextual feature extraction capability.


\vspace{-2pt}
\subsection{Semantic Distillation}
\vspace{-2pt}
The nature of voice conversion can be considered as the decoupling and recombination of semantic information and speaker timbre in speech. As discussed in Expressive-VC~\cite{expressive-vc}, the decoupling process can either be done outside or before the voice conversion model, or rely on the fine-grained design of the voice conversion model itself. 
The DualVC architectures introduced in~\cite{dualvc2,dualvc} follow the popular recognition-synthesis framework, are belong to the former decoupling approach. The pre-trained ASR exhibits excellent speaker-independent semantic information extraction ability and noise robustness. However, integrating this ASR system separately introduces additional complexity to the whole pipeline and causes cascading errors with the VC model. Besides, streaming ASR performs poorly on small chunks, limiting the streaming VC model to further reduce latency. Also, delayed CTC spike distributions and token emission latency existing in streaming ASR leads to semantic information shifting~\cite{streamvoice,trimtail}, causing more potential latency. To this end, we remove the dependency on an external ASR encoder and introduce a pre-trained self-supervised learning (SSL) model for semantic distillation. This model can be omitted during the inference.

Discrete semantic tokens obtained by K-means clustering of SSL features adopted in language model-based generative speech modeling have shown excellent semantic representations with speaker-independent properties~\cite{audiolm, musiclm, speartts}. 
Inspired by this, we perform semantic distillation by using semantic tokens to guide the training of the content encoder. 
Semantic tokens $S = \{s_1, s_2, \ldots, s_{T}\}, s_i \in \{1, 2, \ldots, N\}$ is a sequence of integers extracted from the input audio signal.  Here, $T$ denotes the sequence length, while $N$ denotes the number of clustering centers for K-means.
For an input mel-spectrogram $M \in \mathbb{R}^{T_m\times F}$ with $T_m$ frames and $F$ mel bins, the content encoder extracts the intermediate representation $Z \in \mathbb{R}^{T_m\times D}$ with $D$ dimensions. $Z$ is then downsampled to match the length of semantic tokens and linearly projected to $N$ dimensions to obtain $Z^\prime \in \mathbb{R}^{T\times N}$, then a cross-entropy loss is computed between $Z^\prime$ and $S$ to perform semantic distillation:

\begin{equation}
\mathcal{L}_{CE} = CrossEntropy(Z^\prime, S).
\end{equation}

To further remove residual speaker timbre, $Z^\prime$ is discretized to obtain $ZQ = \{zq_1, zq_2, \ldots, zq_T\}, zq_i \in \{1, 2, \ldots, N\}$, which forms an information bottleneck. The discretization is achieved by Gumbel Softmax~\cite{gumbelsoftmax} which can pass the gradient from the decoder to the encoder. 

Another advantage of using discrete intermediate representation is that it gives the streaming voice conversion model codec-like capabilities. In practice, when deploying the model in a client-server manner, direct audio transfer between the client and the server demands high network bandwidth and can incur notable latency. By using discrete intermediate representations, the bit rate is greatly decreased, significantly reducing network overhead and lowering latency.

\subsection{Language Model for Pseudo Context Genration}

It has been noted that regardless of the structure, all streaming models are inferior to their non-streaming counterparts, with lower intelligibility and poorer speaker similarities. The underlying reason is that context size has a crucial impact on model performance. Since streaming models have no access to future context, it becomes considerably more challenging to achieve optimal performance. As previously discussed, existing approaches try to address this issue by improving the model's capabilities or increasing the amount of information contained in the input features.
In this paper, with an extremely small context size (20 ms), we propose an alternative approach to tackle this problem.

\begin{figure}[ht]
\centering
\includegraphics[scale=0.38]{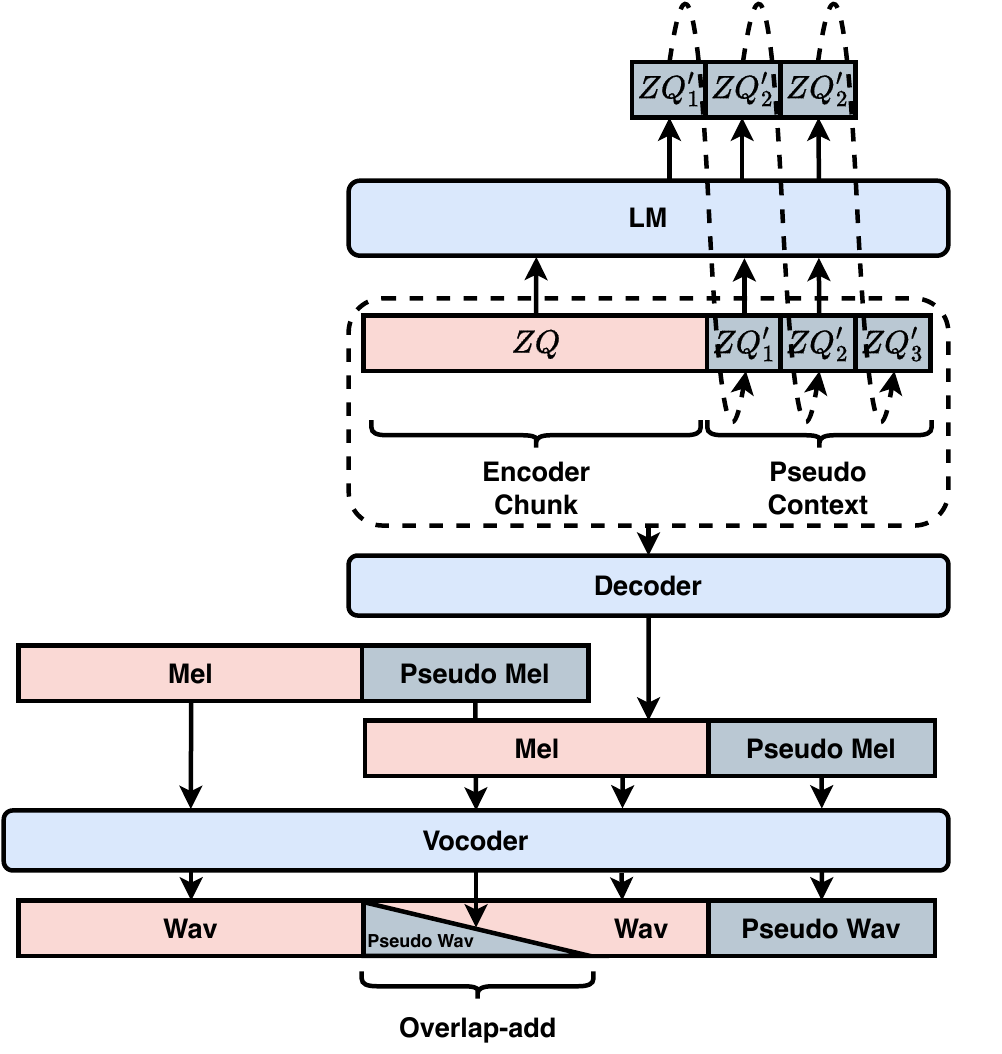}
\vspace{-8pt}
\caption{The inference process of the language model, the decoder and the vocoder.}
\label{fig:lm}
\vspace{-6pt}
\end{figure}
As shown in Fig.\ref{fig:lm}, a language model for pseudo context generation is trained on the discrete intermediate representations $ZQ$ in the typical next-token-prediction manner. During the inference stage, given a chunked $ZQ$ sequence encoded by the encoder, the LM iteratively samples a pseudo context sequence $ZQ^\prime = \{zq^\prime_1, zq^\prime_2, \ldots, zq^\prime_n\}$ from the conditional probability:
\begin{equation}
p_\theta(ZQ^\prime) = \prod^{n}_{i=2}p_\theta(zq^\prime_i | zq^\prime_{i-1}, \cdots, zq^\prime_1, ZQ),
\end{equation}
where $n$ stands for the frames of pseudo-context to be predicted, and $\theta$ is the LM parameters. The concatenation $\{ZQ, ZQ^\prime\}$ is fed to the decoder to synthesize the conversion result.

The process of predicting the pseudo-context is an unconditional continuation process. As the number of predicted frames increases, the features progressively deviate from the ground truth.
But thanks to the DCT strategy adopted to train the conformer-based backbone, the model implicitly assigns decreasing weights to the future context, thus naturally avoiding the intelligibility problem caused by LM prediction errors.

\subsection{Training \& Inference Procedure}
\subsubsection{Training}
The acoustic model, including the encoder and the decoder, is separately trained with the language model. The training objective of the acoustic model includes a reconstruction loss, an HPC loss and a CE loss:

\begin{equation}
\mathcal{L}_{acoustic} = \alpha\mathcal{L}_{rec} + \beta\mathcal{L}_{HPC} + \gamma\mathcal{L}_{CE},
\end{equation}
where $\alpha$, $\beta$, $\gamma$ are weighing terms, which are set to 45, 1, and 10 during training, respectively. The reconstruction loss is an MSE loss calculated between ground-truth mel-spectrogram $M$ and generated one $\hat{M}$:

\begin{equation}
\mathcal{L}_{rec} = MSE(M, \hat{M}).
\end{equation}

Once the acoustic model has converged, we extract the discrete intermediate representations $ZQ$ for training the language model:

\begin{equation}
\mathcal{L}_{LM} = - \sum^{T}_{i=2}\mathrm{log}p_\theta(zq_i|zq_{i-1}, \cdots, zq_{1}).
\end{equation}

\subsubsection{Streaming Inference}
During inference, HPC and Wav2Vec 2.0 are discarded. DualVC 3 can either run in the full mode or in the stand-alone mode with the language model removed. 

\noindent\textbf{Full Mode}
The input mel-spectrogram is first encoded by the content encoder, then discretized to obtain $ZQ$.
$ZQ$ along with pseudo context predicted by the LM and the pre-extracted target speaker embedding are fed to the decoder to generate chunked mel as output.
The pseudo mel generated by the decoder corresponding to the pseudo context is also fed to the vocoder to generate additional pseudo waveforms, enabling overlap-add~\cite{ola} for smoothing between chunks.

\noindent\textbf{Stand-alone Mode}
Only the acoustic model, including the content encoder and the decoder, is preserved in Fig.\ref{fig:model}. With the LM discarded, the computational cost is lowered but also slightly decreases the conversion quality. 


\vspace{-4pt}
\section{Experiments}
\vspace{-4pt}


\subsection{Experimental Setup}
\vspace{-2pt}
\textbf{Dataset}
In the experiments, all testing VC models are trained on an open-source Mandarin corpus AISHELL-3~\cite{aishell3}. This dataset consists of 88,035 utterances spoken by 218 speakers. From these speakers, four males and four females were selected as the target speakers, and 100 speech clips were randomly set aside for evaluation purposes. The selected clips are then converted to the eight target speakers using the proposed model and all comparison models to further perform evaluations. 
All the speech clips are resampled to 16 kHz for VC training. Mel-spectrograms are computed at a frame length of 40 ms and a frame shift of 10 ms.
To increase the amount of data as well as the variety of prosody, we used the open-source tool WavAugment~\footnote{https://github.com/facebookresearch/WavAugment/} to change the tempo.

\noindent\textbf{Implementation Details}
DualVC 3 is composed of 12 Conformer blocks with 256 feature dimensions and 4 self-attention heads. Both the encoder and decoder consist of 6 blocks each. During the DCT training process, there is a 50\% chance of using the full sequence and in the rest of the cases, the chunk size is randomized between 1 (= 10 ms) and 8 (= 80 ms).
The HPC module's future prediction step is set to 6. The speaker embedding is extracted using WeSpeaker toolkit~\cite{wespeaker}. To reconstruct waveform from the converted Mel-spectrograms, we use HiFi-GAN ~\cite{hifigan} with iSTFT upsampling layers~\cite{istftnet} to perform high-fidelity while fast waveform generation. It generates 24 kHz waveforms from 16 kHz spectrograms for better sound quality.
The LM is built using the multi-layer LLaMA~\cite{llama} with unidirectional attention. It comprises 4 layers and 8 heads, with the hidden and intermediate sizes set to 512 and 1024, respectively. The LM predicts 2 frames of pseudo context. The number of K-means clusters is set to 150 to extract semantic tokens.

\noindent\textbf{Comparison Systems}
DualVC 2 and VQMIVC were chosen as baseline models, representing recognition-synthesis-based and SRD-based systems, respectively. The official open-source code~\footnote{https://github.com/Wendison/VQMIVC} was used to reproduce the non-streaming VQMIVC, while the streaming model was achieved by replacing all convolutions with causal convolutions. We compared the combination of two modes for DualVC 3, along with two different chunk sizes (20ms and 160ms).

\begin{table}[]
\caption{Subjective evaluation results in terms of 5-scale naturalness MOS (NMOS) and similarity MOSS (SMOS) with 95\% confidence intervals, along with character error rate (CER).}
\label{tab:mos}
\resizebox{\linewidth}{!}{
\begin{tabular}{l|l|ccc}
\hline
Method                 & Chunk (ms) & NMOS $\uparrow$ & SMOS $\uparrow$ & CER (\%) $\downarrow$ \\ \hline
Ground Truth           & -1         & 4.43$\pm$0.04   & N/A             & 6.2                  \\
DualVC 2~\cite{dualvc2}               & 160        & 3.86$\pm$0.03   & 3.84$\pm$0.06   & 10.3                 \\
VQMIVC~\cite{vqmivc} (non-streaming) & -1         & 2.63$\pm$0.07   & 2.82$\pm$0.09   & 46.6                 \\
VQMIVC~\cite{vqmivc} (streaming)     & 160        & 2.45$\pm$0.02   & 2.76$\pm$0.01   & 57.5                 \\ \hline
DualVC 3 (full)        & 20         & 3.66$\pm$0.02   & 3.76$\pm$0.08   & 14.4                 \\
DualVC 3 (full)        & 160        & 3.73$\pm$0.06   & 3.79$\pm$0.05   & 12.6                 \\
DualVC 3 (stand-alone) & 20         & 3.57$\pm$0.03   & 3.74$\pm$0.07   & 17.8                 \\
DualVC 3 (stand-alone) & 160        & 3.71$\pm$0.05   & 3.71$\pm$0.03   & 13.2                 \\ \hline
\end{tabular}
}
\vspace{-6pt}
\end{table}

\subsection{Subjective Evaluation}
We conduct Mean Opinion Score (MOS) tests to evaluate the naturalness and speaker similarity of comparison models. The naturalness metric mainly considers intelligibility, prosody, and sound quality. A higher naturalness MOS score indicates the converted speech sounds more human-like. The similarity test uses the target speaker's real recording as the reference to evaluate the timbre similarity between real and converted recordings. In both MOS tests, there are 30 listeners participated. 

\noindent\textbf{Speech Naturalness}
The naturalness MOS results presented in Table~\ref{tab:mos} indicate that DualVC 3 achieves fair performance with only a 20ms chunk size. With the inclusion of extra pseudo context, the model shows significant performance improvement. However, as the chunk size increases, the benefit of pseudo context decreases. VQMIVC performs poorly on the mandarin corpus, with further degradation of effects under streaming. This demonstrates that SRD-based models hardly achieve good results in streaming scenarios.

\noindent\textbf{Speaker Similarity}
DualVC 3 achieves high speaker similarity, significantly outperforming the SRD-based VQMIVC and approaching the performance of DualVC 2. This validates the effectiveness of semantic distillation. The SMOS scores were consistent across different model configurations of DualVC 3, possibly indicating that timbre, as a global feature, is relatively unaffected by context size.

\subsection{Objective Evaluation}
\noindent\textbf{Intelligibility Evaluation}
We employ a conformer-based ASR model pre-trained on WenetSpeech~\cite{wenetspeech} to transcribe the source and converted speech. To ensure the accuracy of our results, we conduct testing on a larger dataset comprising 500 samples. The Character Error Rate (CER) is also detailed in Table~\ref{tab:mos}. For the source speech, we observed a CER of 6.2\%. The CER of DualVC 3 demonstrates close adherence to the NMOS score,  with lower CER in large context or full mode with pseudo context.

\begin{figure}
  \centering
  \includegraphics[width=0.40\textwidth]{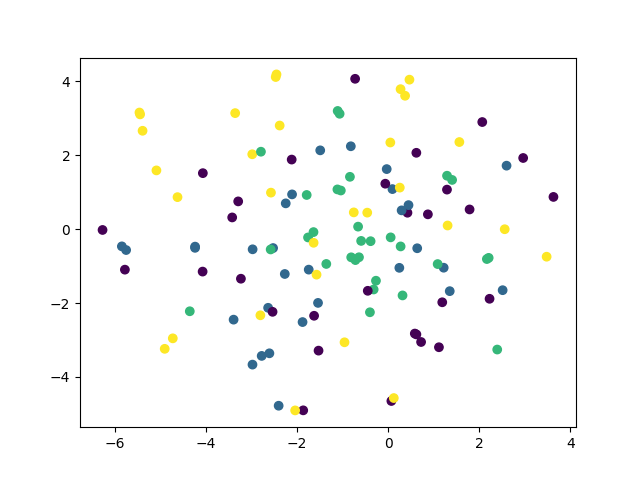}
  \vspace{-6pt}
  \caption{Encoder output visualization using t-SNE. Each color represents a speaker.}
  \label{fig:tsne}
  \vspace{-6pt}
\end{figure}

\noindent\textbf{Visualization of Encoder Output}
To demonstrate the decoupling ability of the proposed semantic distillation approach, we visualize the encoder outputs by t-SNE~\cite{tsne}. Thirty utterances from 4 source speakers are selected. As shown in Fig~\ref{fig:tsne}, the encoder outputs are projected to 2D by t-SNE, with each color representing a speaker. With the guidance of discrete semantic tokens, the encoder successfully extracts speaker-independent semantic information.

\noindent\textbf{Computational Efficiency Evaluation}
As illustrated in Table~\ref{tab:performance}, the overall latency is 43.58 ms and 55.94 ms for stand-alone mode and full mode respectively, which consists of model inference latency, chunk-waiting latency (20ms), and lookahead latency (20ms). Note that with more pseudo context generated by the LM in full mode, the RTF and latency are higher for the AM and the vocoder.
\begin{table}[]
\centering
 \caption{Real-time \& Computation Metrics of DualVC 3 tested on single core i5 10210U.}
 \vspace{-6pt}
 \label{tab:performance}
\resizebox{\linewidth}{!}{
\begin{tabular}{l|ccc}
\hline
                        & RTF   & Latency (ms)      & Params (M) \\ \hline
Full mode               & 0.797 & 15.94+20+20=55.94 & 22.7       \\
\hspace{1em}AM (w/ 2 pseudo ctx) & 0.201 & 4.02              & 10.9       \\
\hspace{1em}Vocoder (w/ 2 pseudo ctx)  & 0.086 & 1.72              & 1.2        \\
\hspace{1em}LM                      & 0.510 & 10.20             & 10.6       \\ \hline
Stand-alone mode        & 0.181 & 3.58+20+20=43.58  & 12.1       \\ 
\hspace{1em}AM (w/o pseudo ctx)               & 0.134 & 2.68              & 10.9       \\
\hspace{1em}Vocoder (w/o pseudo ctx)                & 0.047 & 0.90              & 1.2        \\\hline
\end{tabular}
}
\vspace{-12pt}
\end{table}

\section{Conclusions}
In this paper, we propose an end-to-end streaming voice conversion model DualVC 3 with latency of only 50 ms. With semantic tokens as guidance, the content encoder successfully extracts speaker-independent semantic information without the need for ASR. A language model is adopted to generate pseudo context for the decoder to improve conversion quality. Experiments show that DualVC 3 is more than 3 times faster than DualVC 2, with comparable performance.
\newpage

\bibliographystyle{IEEEtran}
\bibliography{mybib}

\begin{thebibliography}{10}
\providecommand{\url}[1]{#1}
\csname url@samestyle\endcsname
\providecommand{\newblock}{\relax}
\providecommand{\bibinfo}[2]{#2}
\providecommand{\BIBentrySTDinterwordspacing}{\spaceskip=0pt\relax}
\providecommand{\BIBentryALTinterwordstretchfactor}{4}
\providecommand{\BIBentryALTinterwordspacing}{\spaceskip=\fontdimen2\font plus
\BIBentryALTinterwordstretchfactor\fontdimen3\font minus \fontdimen4\font\relax}
\providecommand{\BIBforeignlanguage}[2]{{%
\expandafter\ifx\csname l@#1\endcsname\relax
\typeout{** WARNING: IEEEtran.bst: No hyphenation pattern has been}%
\typeout{** loaded for the language `#1'. Using the pattern for}%
\typeout{** the default language instead.}%
\else
\language=\csname l@#1\endcsname
\fi
#2}}
\providecommand{\BIBdecl}{\relax}
\BIBdecl

\bibitem{vcoverview}
B.~Sisman, J.~Yamagishi, S.~King, and H.~Li, ``An overview of voice conversion and its challenges: From statistical modeling to deep learning,'' \emph{{IEEE} {ACM} Trans. Audio Speech Lang. Process.}, vol.~29, pp. 132--157, 2021.

\bibitem{bgmvc}
J.~Yao, Y.~Lei, Q.~Wang, P.~Guo, Z.~Ning, L.~Xie, H.~Li, J.~Liu, and D.~Xie, ``Preserving background sound in noise-robust voice conversion via multi-task learning,'' in \emph{Proc. ICASSP}.\hskip 1em plus 0.5em minus 0.4em\relax {IEEE}, 2023, pp. 1--5.

\bibitem{expressive-vc}
Z.~Ning, Q.~Xie, P.~Zhu, Z.~Wang, L.~Xue, J.~Yao, L.~Xie, and M.~Bi, ``Expressive-vc: Highly expressive voice conversion with attention fusion of bottleneck and perturbation features,'' in \emph{Proc. ICASSP}.\hskip 1em plus 0.5em minus 0.4em\relax {IEEE}, 2023, pp. 1--5.

\bibitem{anonymization}
J.~Yao, Q.~Wang, Y.~Lei, P.~Guo, L.~Xie, N.~Wang, and J.~Liu, ``Distinguishable speaker anonymization based on formant and fundamental frequency scaling,'' in \emph{Proc. ICASSP}.\hskip 1em plus 0.5em minus 0.4em\relax {IEEE}, 2023, pp. 1--5.

\bibitem{ELECTROLARYNGEAL}
K.~Kobayashi, T.~Hayashi, and T.~Toda, ``Low-latency electrolaryngeal speech enhancement based on fastspeech2-based voice conversion and self-supervised speech representation,'' in \emph{Proc. ICASSP}, 2023, pp. 1--5.

\bibitem{freevc}
J.~Li, W.~Tu, and L.~Xiao, ``Freevc: Towards high-quality text-free one-shot voice conversion,'' in \emph{Proc. ICASSP}.\hskip 1em plus 0.5em minus 0.4em\relax {IEEE}, 2023, pp. 1--5.

\bibitem{acevc}
S.~Hussain, P.~Neekhara, J.~Huang, J.~Li, and B.~Ginsburg, ``{ACE-VC:} adaptive and controllable voice conversion using explicitly disentangled self-supervised speech representations,'' in \emph{Proc. ICASSP}.\hskip 1em plus 0.5em minus 0.4em\relax {IEEE}, 2023, pp. 1--5.

\bibitem{vqmivc}
D.~Wang, L.~Deng, Y.~T. Yeung, X.~Chen, X.~Liu, and H.~Meng, ``{VQMIVC:} vector quantization and mutual information-based unsupervised speech representation disentanglement for one-shot voice conversion,'' in \emph{Proc. INTERSPEECH}.\hskip 1em plus 0.5em minus 0.4em\relax {ISCA}, 2021, pp. 1344--1348.

\bibitem{stylevc}
Z.~Wang, X.~Zhou, F.~Yang, T.~Li, H.~Du, L.~Xie, W.~Gan, H.~Chen, and H.~Li, ``Enriching source style transfer in recognition-synthesis based non-parallel voice conversion,'' in \emph{Proc. INTERSPEECH}.\hskip 1em plus 0.5em minus 0.4em\relax {ISCA}, 2021, pp. 831--835.

\bibitem{ibfvc}
Y.~Chen, M.~Tu, T.~Li, X.~Li, Q.~Kong, J.~Li, Z.~Wang, Q.~Tian, Y.~Wang, and Y.~Wang, ``Streaming voice conversion via intermediate bottleneck features and non-streaming teacher guidance,'' in \emph{Proc. ICASSP}.\hskip 1em plus 0.5em minus 0.4em\relax {IEEE}, 2023, pp. 1--5.

\bibitem{dualvc}
Z.~Ning, Y.~Jiang, P.~Zhu, J.~Yao, S.~Wang, L.~Xie, and M.~Bi, ``Dualvc: Dual-mode voice conversion using intra-model knowledge distillation and hybrid predictive coding,'' in \emph{Proc. INTERSPEECH}.\hskip 1em plus 0.5em minus 0.4em\relax {ISCA}, 2023, pp. 2063--2067.

\bibitem{dualvc2}
Z.~Ning, Y.~Jiang, P.~Zhu, S.~Wang, J.~Yao, L.~Xie, and M.~Bi, ``Dualvc 2: Dynamic masked convolution for unified streaming and non-streaming voice conversion,'' in \emph{Proc. ICASSP}.\hskip 1em plus 0.5em minus 0.4em\relax {IEEE}, 2024, pp. 1--5.

\bibitem{DBLP:conf/icassp/HayashiKT22}
T.~Hayashi, K.~Kobayashi, and T.~Toda, ``An investigation of streaming non-autoregressive sequence-to-sequence voice conversion,'' in \emph{Proc. ICASSP 2022}.\hskip 1em plus 0.5em minus 0.4em\relax {IEEE}, 2022, pp. 6802--6806.

\bibitem{cuside}
K.~An, H.~Zheng, Z.~Ou, H.~Xiang, K.~Ding, and G.~Wan, ``{CUSIDE:} chunking, simulating future context and decoding for streaming {ASR},'' in \emph{Proc. INTERSPEECH}.\hskip 1em plus 0.5em minus 0.4em\relax {ISCA}, 2022, pp. 2103--2107.

\bibitem{streamvoice}
Z.~Wang, Y.~Chen, X.~Wang, Z.~Chen, L.~Xie, Y.~Wang, and Y.~Wang, ``Streamvoice: Streamable context-aware language modeling for real-time zero-shot voice conversion,'' \emph{CoRR}, vol. abs/2401.11053, 2024.

\bibitem{streamingvqmivc}
H.~Yang, L.~Deng, Y.~T. Yeung, N.~Zheng, and Y.~Xu, ``Streamable speech representation disentanglement and multi-level prosody modeling for live one-shot voice conversion,'' in \emph{Proc. INTERSPEECH 2022}.\hskip 1em plus 0.5em minus 0.4em\relax {ISCA}, pp. 2578--2582.

\bibitem{fasts2svc}
H.~Kameoka, K.~Tanaka, and T.~Kaneko, ``Fasts2s-vc: Streaming non-autoregressive sequence-to-sequence voice conversion,'' \emph{CoRR}, vol. abs/2104.06900, 2021.

\bibitem{w2v2}
A.~Baevski, Y.~Zhou, A.~Mohamed, and M.~Auli, ``wav2vec 2.0: {A} framework for self-supervised learning of speech representations,'' in \emph{Proc. NeurIPS}, 2020, pp. 12\,449--12\,460.

\bibitem{audiolm}
Z.~Borsos, R.~Marinier, D.~Vincent, E.~Kharitonov, O.~Pietquin, M.~Sharifi, D.~Roblek, O.~Teboul, D.~Grangier, M.~Tagliasacchi, and N.~Zeghidour, ``Audiolm: {A} language modeling approach to audio generation,'' \emph{{IEEE} {ACM} Trans. Audio Speech Lang. Process.}, vol.~31, pp. 2523--2533, 2023.

\bibitem{musiclm}
A.~Agostinelli, T.~I. Denk, Z.~Borsos, J.~H. Engel, M.~Verzetti, A.~Caillon, Q.~Huang, A.~Jansen, A.~Roberts, M.~Tagliasacchi, M.~Sharifi, N.~Zeghidour, and C.~H. Frank, ``Musiclm: Generating music from text,'' \emph{CoRR}, vol. abs/2301.11325, 2023.

\bibitem{wenet}
Z.~Yao, D.~Wu, X.~Wang, B.~Zhang, F.~Yu, C.~Yang, Z.~Peng, X.~Chen, L.~Xie, and X.~Lei, ``Wenet: Production oriented streaming and non-streaming end-to-end speech recognition toolkit,'' in \emph{Proc. INTERSPEECH}.\hskip 1em plus 0.5em minus 0.4em\relax {ISCA}, 2021, pp. 4054--4058.

\bibitem{cpc}
A.~van~den Oord, Y.~Li, and O.~Vinyals, ``Representation learning with contrastive predictive coding,'' \emph{CoRR}, vol. abs/1807.03748, 2018.

\bibitem{apc}
Y.~Chung, W.~Hsu, H.~Tang, and J.~R. Glass, ``An unsupervised autoregressive model for speech representation learning,'' in \emph{Proc. INTERSPEECH}.\hskip 1em plus 0.5em minus 0.4em\relax {ISCA}, 2019, pp. 146--150.

\bibitem{trimtail}
X.~Song, D.~Wu, Z.~Wu, B.~Zhang, Y.~Zhang, Z.~Peng, W.~Li, F.~Pan, and C.~Zhu, ``Trimtail: Low-latency streaming {ASR} with simple but effective spectrogram-level length penalty,'' in \emph{Proc. ICASSP}.\hskip 1em plus 0.5em minus 0.4em\relax {IEEE}, 2023, pp. 1--5.

\bibitem{speartts}
E.~Kharitonov, D.~Vincent, Z.~Borsos, R.~Marinier, S.~Girgin, O.~Pietquin, M.~Sharifi, M.~Tagliasacchi, and N.~Zeghidour, ``Speak, read and prompt: High-fidelity text-to-speech with minimal supervision,'' \emph{CoRR}, vol. abs/2302.03540, 2023.

\bibitem{gumbelsoftmax}
E.~Jang, S.~Gu, and B.~Poole, ``Categorical reparameterization with gumbel-softmax,'' in \emph{Proc. ICLR 2017}.

\bibitem{ola}
Z.~Wang and S.~Watanabe, ``Improving frame-online neural speech enhancement with overlapped-frame prediction,'' \emph{{IEEE} Signal Process. Lett.}, vol.~29, pp. 1422--1426, 2022.

\bibitem{aishell3}
Y.~Shi, H.~Bu, X.~Xu, S.~Zhang, and M.~Li, ``{AISHELL-3:} {A} multi-speaker mandarin {TTS} corpus,'' in \emph{Proc. INTERSPEECH}.\hskip 1em plus 0.5em minus 0.4em\relax {ISCA}, 2021, pp. 2756--2760.

\bibitem{wespeaker}
H.~Wang, C.~Liang, S.~Wang, Z.~Chen, B.~Zhang, X.~Xiang, Y.~Deng, and Y.~Qian, ``Wespeaker: {A} research and production oriented speaker embedding learning toolkit,'' in \emph{Proc. ICASSP}.\hskip 1em plus 0.5em minus 0.4em\relax {IEEE}, 2023, pp. 1--5.

\bibitem{hifigan}
J.~Su, Z.~Jin, and A.~Finkelstein, ``Hifi-gan: High-fidelity denoising and dereverberation based on speech deep features in adversarial networks,'' in \emph{Proc. INTERSPEECH}.\hskip 1em plus 0.5em minus 0.4em\relax {ISCA}, 2020, pp. 4506--4510.

\bibitem{istftnet}
T.~Kaneko, K.~Tanaka, H.~Kameoka, and S.~Seki, ``{ISTFTNET:} fast and lightweight mel-spectrogram vocoder incorporating inverse short-time fourier transform,'' in \emph{Proc. ICASSP}.\hskip 1em plus 0.5em minus 0.4em\relax {IEEE}, 2022, pp. 6207--6211.

\bibitem{llama}
H.~Touvron, T.~Lavril, G.~Izacard, X.~Martinet, M.~Lachaux, T.~Lacroix, B.~Rozi{\`{e}}re, N.~Goyal, E.~Hambro, F.~Azhar, A.~Rodriguez, A.~Joulin, E.~Grave, and G.~Lample, ``Llama: Open and efficient foundation language models,'' \emph{CoRR}, vol. abs/2302.13971, 2023.

\bibitem{wenetspeech}
B.~Zhang, H.~Lv, P.~Guo, Q.~Shao, C.~Yang, L.~Xie, X.~Xu, H.~Bu, X.~Chen, C.~Zeng, D.~Wu, and Z.~Peng, ``{WENETSPEECH:} {A} 10000+ hours multi-domain mandarin corpus for speech recognition,'' in \emph{Proc. ICASSP}.\hskip 1em plus 0.5em minus 0.4em\relax {IEEE}, 2022, pp. 6182--6186.

\bibitem{tsne}
L.~Van~der Maaten and G.~Hinton, ``Visualizing data using t-sne.'' \emph{Journal of machine learning research}, vol.~9, no.~11, 2008.

\end{thebibliography}

\end{document}